\documentclass[aps,pre,twocolumn,showpacs,superscriptaddress]{revtex4}

\usepackage{bm}
\usepackage{graphicx}
\usepackage{amsmath}
\usepackage{amssymb}
\usepackage[dvips]{color}

\begin{document}
\def\be{\begin{equation}}
\def\ee{\end{equation}}

\def\bfi{\begin{figure}}
\def\efi{\end{figure}}
\def\bea{\begin{eqnarray}}
\def\eea{\end{eqnarray}}

\title{Rattler-induced aging dynamics in jammed granular systems}

\author{F. Giacco}
\affiliation{
Dept. of Mathematics and Physics, University of Campania ``Luigi Vanvitelli'', Caserta, Italy}

\author{L. de Arcangelis}
\affiliation{
Dept. of Industrial and Information Engineering, University of Campania ``Luigi Vanvitelli'', Aversa (CE), Italy}

\author{M. Pica Ciamarra}
\affiliation{
Division of Physics and Applied Physics, School of
Physical and Mathematical Sciences, Nanyang Technological University, Singapore}
\affiliation{CNR--SPIN, Dept. of Physics, University of Naples ``Federico II'', 
Naples, Italy}

\author{E. Lippiello} 
\affiliation{
Dept. of Mathematics and Physics, University of Campania ``Luigi Vanvitelli'', Caserta, Italy}

\date{\today}

\begin{abstract}
Granular materials jam when developing a network of contact forces able to resist the applied stresses.
Through numerical simulations of the dynamics of the jamming process, we show that the jamming transition
does not occur when the kinetic energy vanishes. Rather, as the system jams, the kinetic energy becomes
dominated by rattlers particles, that scatter withing their cages. The relaxation of the kinetic energy
in the jammed configuration exhibits a double power-law decay, which we interpret in terms of the interplay
between backbone and rattlers particles.
\end{abstract}


\maketitle

\section{Introduction}
The solid to liquid transition of granular systems, besides controlling
many natural phenomena such as earthquakes or landslides,
is of great theoretical interest since it is related to 
an out-of-equilibrium phase transition~\cite{corwin2015,maksephyschemestry,johnsonjia2005,johnsonsava2008,johnson2012,vanderelst2012}
and to the physics of glass forming systems.
The control parameters that drive this transition are the density/pressure, the shear strain/stress~\cite{liu_nonlinear_1998},
as well as the frictional properties of the grains~\cite{song_phase_2008,pica_ciamarra_jamming_2011}. 
This transition has been extensively investigated in the static limit, where the grain dynamics is either absent or negligible.
For instance, the earliest investigation
of the transition~\cite{ohern_jamming_2003} considered the evolution of the properties
of jammed packings of frictionless grains as the density approaches the critical value
and a number of experimental and numerical following works have been conducted in the same spirit~\cite{liu_jamming_2010,van_hecke_jamming_2009}.
Similarly, the shear induced transition from the jammed to the flowing state has been mainly investigated in the limit
of quasistatic deformations. In numerical simulations, this limit is realized by minimizing the energy after 
every infinitesimal increase of the shear strain~\cite{arevalo_size_2014}, which makes inertial effects irrelevant.
While the quasistatic approximation is reasonable when describing the transition from the solid
to the fluid state, since in the solid phase particle motion is negligible, it must be relaxed to describe
the transition from the flowing to the jammed state. Indeed, inertial events make 
the jamming/unjamming transition hysteretic~\cite{PhysRevLett.103.235701} by affecting
the location of the transition threshold.

In this paper we numerically investigate the transition from the unjammed to the jammed state
in a model system exhibiting stick-slip motion. The model consists
of a collection of grains confined in between two rigid plates at constant pressure. The bottom plate is fixed, while
the top one is driven through a spring mechanism, as in spring-block models, which leads to a stick-slip motion.
This and similar models have been investigated by a number of authors in both experiments and 
simulations~\cite{prl2010,epl2011,modphys2009,petri_europjournal,giaccoprl2015,giaccopre,prl2010,GRL:GRL50813,griffa2013,3ddegriffa2014,ciamarra_role_2012},
that focused on the identification of slip precursors, on the study of the response to external perturbations~\cite{giaccoprl2015,melosh79,melosh96,xia2011pre,xiamarone2013,capozzaprl}
and on the characterization of the slip size distribution, that has been
shown to be affected by inertial effects~\cite{PhysRevLett.109.105703}.
In this model, a slip starts when the granular system becomes 
unable to sustain the shear stress exerted by the top plate, 
and it ends as soon as it becomes able to balance again the shear stress whose value
has decreased because of the slip.
In this jammed configuration, the velocity of the top plate vanishes, and a network of contact forces
between the grains counter-balance the applied stress~\cite{prl2010}. 
Here we show that, surprisingly, the kinetic energy of the systems is not zero when the system jams.
Rather, the kinetic energy never vanishes, but decreases in time as a power law,
with a crossover between a slower decay at short times and a faster decay at 
long times. 
We show that this behavior originates from coupled structural and dynamical 
heterogeneities, due to the coexistence of 
particles forming the sustaining backbone 
and rattlers free to move in cages formed by the backbone particles.  
Indeed, the first slower relaxation regime is affected by both the backbone 
particles and the rattlers, while the second one is dominated by the rattlers. The
crossover between the two regimes is detected at a 
characteristic time $\tau_c$, depending on external constraints. 

%\textcolor{blue}{
%potremmo dire qualcosa sulle oscillazioni.
%potremmo fare una connessione con strictly jammed e jammed.
%The backbone particles during the relaxation process
%perform collective vibrations tracing an elliptic trajectory whose orientation
%affects the nucleation of instabilities. Indeed, only before the occurrence of a 
%slip  the vibrations become perpendicular to the confining plates, 
%reducing the confining pressure and promoting failure, according to the 
%%acoustic fluidization (AF) scenario 
%\cite{giaccoprl2015,melosh79,melosh96,xia2011pre,xiamarone2013,capozzaprl}. 
%}
%\begin{figure}[!t]
%\begin{center}
%\includegraphics[scale=1]{multiplot2.eps}
%\end{center}
%\caption{(Color online)
%(a) Our model system consists of a collection of grains (gray particles) confined in between two rough plates at constant pressure. 
%The top plate is driven through a spring whose free end moves with a constant velocity $V_d > 0$.
%Panels (b) and (c) illustrate the time evolution of the position and of the velocity of top plate, in a short time interval in which two slips occurs.
%Panel (d) illustrates that in the undriven replicas the top plate oscillates
%and comes to a rest in short time.
%} 
%\label{fig:system}
%\end{figure}
\begin{figure}[!t]
\begin{center}
\includegraphics[scale=0.22]{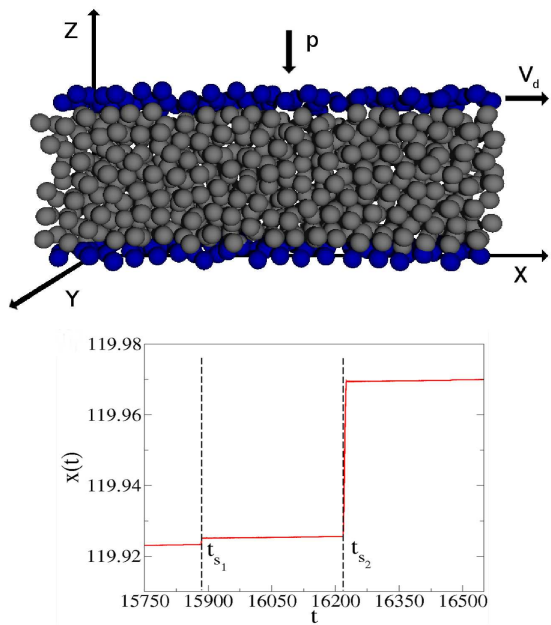}
%\vspace{100pt}
%\includegraphics[scale=0.22]{fig1b.eps}
\end{center}
\caption{(Color online)
(Upper panel) Our model system consists of a collection of grains (gray particles) confined in between two rough plates at constant pressure. 
The top plate is driven through a spring whose free end moves with a constant velocity $V_d > 0$.
(Lower panel) The time evolution of the top plate position during a short time interval in which two slips occurs.
} 
\label{fig:system}
\end{figure}
\section{Model}
We perform three-dimensional molecular dynamics simulations of the model illustrated in Fig.~\ref{fig:system}a,
consisting of $N$ monodisperse spheres of unitary mass $m$ and diameter $d$, enclosed 
between two rigid 
rough plates of dimension $L_x \times L_y$. Each plate is made of $L_x L_y/d^2$ grains, placed in 
random positions in the $xy$-plane. The $z$ position of these grains is randomly shifted to make 
the plates corrugated. The relative positions of the particles 
belonging to the plates are fixed in order to make the plates rigid. The top plate is subject to a
constant pressure $p$ applied along the $z$ direction and attached to a spring 
of elastic constant $k_m$, 
whose free end moves with constant velocity $V_d$ in the $x$
direction. We employ a contact force model described in~\cite{silbert} where 
particles interact along the normal direction via the standard spring-dashpot 
model with a restitution coefficient $e$. We measure the mass in units of $m$, the length in units of $d$ and time in 
units of $\sqrt{m/k_{m}}$. 
All model parameters have been chosen 
according to Ref.~\cite{prl2010,epl2011} in order to have long stick phases 
interrupted by rapid plate displacements, i.e the slips, namely: $N = 1000$, $L_x \times 
L_y = 20\,d \times 5\,d$, $p= {k_m/d}$, $e=0.88$, $V_d=0.01\, d/\sqrt{m/k_{m}}$.
Due to the constant pressure condition, the size of the system along the vertical
direction is not fixed, but slightly fluctuates around $L_z \simeq 10$.
The majority of slips $S_i$ involve small displacements of the top 
plate, i.e. of the order of a small fraction of a grain diameter $S_i \ll d$, 
however also slips involving the whole system $S_i \sim L_x$ are observed~\cite{prl2010,epl2011}.
The temporal integration step of the equations of motion is $5\times 10^{-3}\,\sqrt{m/k_{m}}$.

\begin{figure}[!t]
\begin{center}
\includegraphics[scale=0.3]{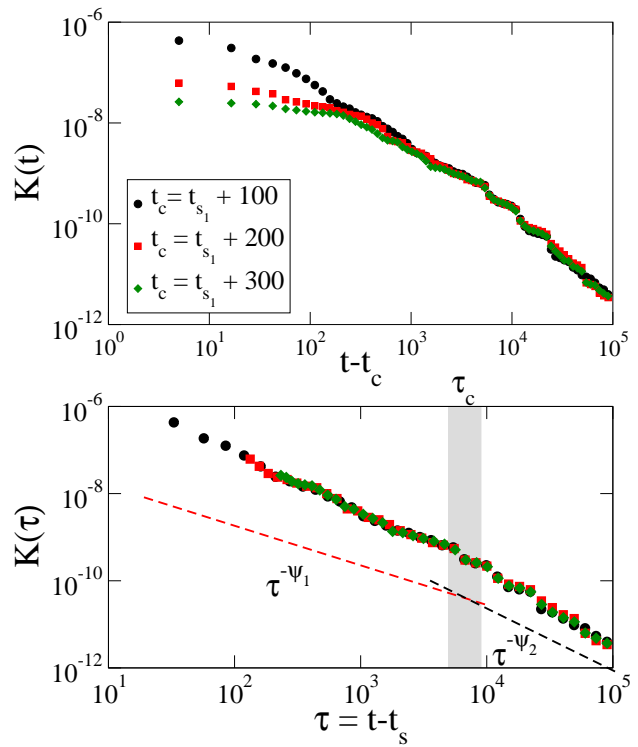} 
\vskip 0.5cm
\end{center}
\caption{(Color online) Kinetic energy relaxation of several replicas of the 
system with different temporal distances from two slip occurrence times 
$t_{s_1}$ 
and $t_{s_2}$. (Upper panel) The short time behavior of the energy $K(t)$ 
clearly depends on the value of $t-t_s$, while it  is mainly 
$(t-t_s)$-independent 
at large times.  (Lower panel) 
Kinetic energy is plotted as a function 
of $\tau=t-t_s$.  Data obtained for different $t-t_s$ collapse onto a 
single 
curve, exhibiting two power law regimes.} 
\label{fig1}
\end{figure}

\section{Results}
\subsection{Relaxation after a slip}
During a slip, the system dissipates energy and the applied stress decreases. 
When the slip is over, the system should be found,
in principle, in a jammed state in which the top plate velocity is zero. 
This only occurs, however, in the limit $V_d \to 0$, 
since for finite $V_d$ in the jammed state the top plate
exhibits a creep motion as the system continuously adapts to the increasing shear stress. 
%We show that this is the case in Fig.~\ref{fig:system}. 
Figure 1b illustrates the time evolution of the position of the top plate of our system,
in a temporal window where two  slips occur at time $t_{s_i}$, $i = 1,2$. 
Specifically, considering that a slip has a finite duration,
here we define $t_s$ as the time at which the slip ends~\cite{prl2010}, 
so that $t=t_s$ corresponds to the onset of the stick phase. 
We define $t_s$ as the time at which the velocity of the plate becomes smaller than a given threshold, $10^{-2}$.
For $t > t_s$ the velocity of the top decreases in time while
exhibiting a jerking motion.
This is a consequence of the creep motion that sets in due to the increasing shear stress controlled by the driving velocity $V_d > 0$.
To investigate the relaxation dynamics following a slip without being affected 
by the creep motion and by subsequent slips,
we investigate the jamming dynamics of replicas of the system that evolves with zero driving velocity, $V_d = 0$. 
We create replicas at many different times $t_c$ after the slip time $t_s$.
During the relaxation dynamics the top plate of the replica performs damped oscillations and rapidly reaches  a rest position, confirming the absence of creep motion.

Since the velocity of the top plate of the replica evolving with $V_d = 0$ vanishes, 
one might expect the system to reach a jammed state in which there is no particle motion.
Conversely, the investigation of the time evolution of the kinetic energy of the replicas
(Fig.~\ref{fig1}) reveals that this is not the case. 
Indeed, even if the system is in a macroscopic jammed state, the 
decay of the kinetic energy, measured from the time of replica creation, $t-t_c$,
is consistent with a 
power law indicating that the stationary condition $K=0$ is never 
reached. The system, therefore, never attains a state of mechanical 
equilibrium. Fig.~\ref{fig1} (upper panel) shows that the kinetic energy depends
on the time $t_c$ at which the replica is made, and decreases as $t_c$ increases.
Conversely, data collapse when plotted 
as function of the time since the last slip $\tau = t-t_s$, as in the  lower panel of Fig.~\ref{fig1}.
Two different 
regimes in the temporal decay of the kinetic energy are detected, 
with a crossover occurring at $\tau_c \simeq 7000$.
For $\tau < \tau_c$, the kinetic energy decreases as $K \sim \tau^{-\psi_1}$,
while for $\tau > \tau_c$ it decreases as $K \sim \tau^{-\psi_2}$,
Averaging over 10 different slips, we estimate $\psi_1=1.3\pm 0.1$ and $\psi_2=1.8\pm 0.1$.
Combining the two scaling regimes, we expect
\be
K(\tau)=A \tau^{-\psi_1} \left (\frac{\tau}{\tau_c}+1 \right )^{-\psi_2+\psi_1}.
\label{eq1}
\ee 
As we will describe later, this equation holds for any $\tau$ larger than a microscopic time $t_0$ and 
is observed for slips of different sizes with the value of $\tau_c$ and $A$ depending on the 
particular slip. 
\begin{figure}[!t]
\begin{center}
\includegraphics[scale=0.32]{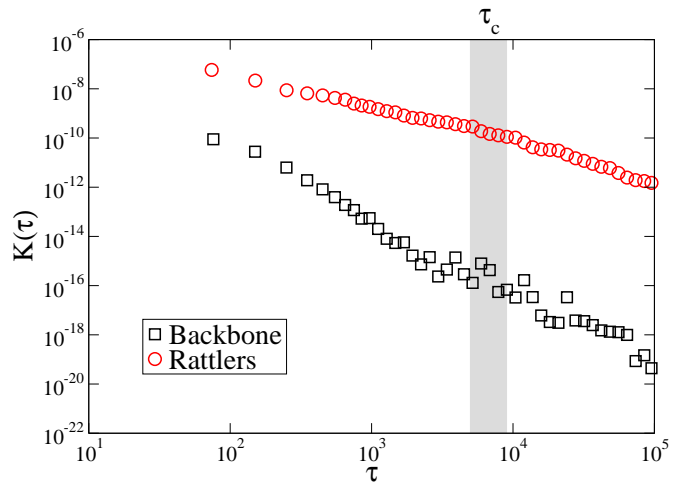}
\end{center}
\caption{\label{fig:erattlers}
(Color online)
Kinetic energy evaluated separately for 
rattlers and backbone particles, for one of the replicas of the system.
Rattlers dominate the overall kinetic energy. The backbone energy exhibits
a crossover from a faster to a slower decay at time $\tau_c$.}
\end{figure}

We next show that the observed relaxation dynamics results from the interplay between rattlers and 
backbone. 
Indeed, in our system we observe that during the relaxation dynamics a small fraction of particles, 
less than $10\%$, are located inside cages and are not in permanent contact with 
other particles. 
These are the rattlers, each of which moves in a cage of volume $\sim (1+10^{-3})d^3$.
In Fig.~\ref{fig:erattlers}  we evaluate the separate contribution 
to the kinetic energy from the backbone particles and 
the rattlers, for a given replica, but analogous results are found for all replicas. 
The figure clarifies that, despite their small number, 
rattlers dominate the total kinetic energy of the system.
This result allows to rationalize why the kinetic energy of
a replica made a time $t_c$ only depends on the time elapsed since the 
preceding slip, as in the lower panel of Fig.~\ref{fig1}: Being the kinetic
energy dominated by the rattlers, this is not influenced by the creep motion,
and it is therefore insensitive to the time $t_c$ at which the creep motion is suppressed
by setting the driving velocity to zero.
Fig.~\ref{fig:erattlers} also shows that the $\tau$ dependence of the kinetic energy
of the backbone exhibits a clear change of behavior at the crossover time $\tau_c$,
becoming much slower for $\tau > \tau_c$. This suggests a physical interpretation
in which the crossover in the decay of the kinetic energy relates to the stiffening
of the backbone, we describe in the following.

\subsection{Physical origin of the crossover}
\begin{figure}[!!t]
\begin{center}
\includegraphics*[scale=0.33]{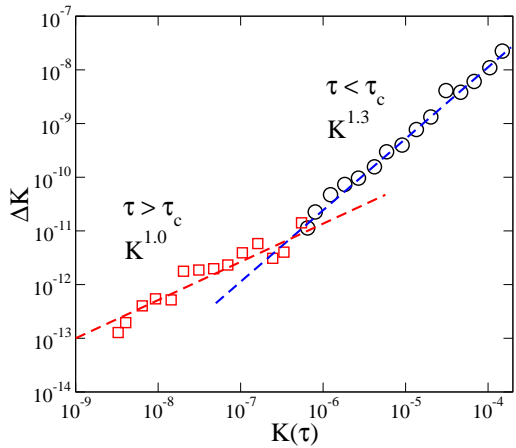}
\end{center}
\caption{Kinetic energy dissipated in a small time interval $d\tau$, 
as a function of the initial kinetic energy $K(\tau)$. The dissipated energy
scales as $K$ and as $K^{1.3}$, 
respectively for $\tau > \tau_c$ and for $\tau <\tau_c$.} 
\label{fig2}
\end{figure}

To rationalize the crossover in the decay of the kinetic energy, Eq.~\ref{eq1}, 
we consider that energy is dissipated though inter-particle collisions. 
Let's suppose that rattlers colliding with the backbone with typical kinetic energy $K$ lose an amount of energy scaling 
as $\Delta K\sim K^{1+\epsilon}$.
Thus, the typical energy variation $dK$, in a 
time interval $d \tau$, is proportional to the average number of collisions 
$n_{d\tau}\propto K^{1/2}$ in $d\tau$ times the average energy dissipated in 
each collision $\Delta K$, $dK/d\tau \propto -n_{d\tau} \Delta K$. 
This leads to $dK K^{-3/2-\epsilon} \propto - d\tau $, 
so that $K \sim \tau^\psi$, with $\psi=\frac{2}{1+2\epsilon}$.
Accordingly, the crossover in $\psi$ occurring at time $\tau_c$
should correspond to a crossover in $\epsilon$.
We have explicitly checked this prediction in numerical simulations,
investigating the kinetic energy lost $\Delta K =K(\tau+d \tau)-K(\tau)$ 
as function of $K(\tau)$, for different values of $\tau$
and a fixed small value of $d\tau$. 
Fig.~\ref{fig2} shows that $\Delta K$ does actually exhibit a crossover
at $\tau_c$, scaling as $\Delta K \simeq K^{1+\epsilon}$ 
with $\epsilon \simeq 0.3$ for $\tau < \tau_c$, and
$\epsilon\simeq 0$ for $\tau > \tau_c$.
These values of $\epsilon$ are consistent with the expected
values of $\psi=\frac{2}{1+2\epsilon}$.

Physically, the crossover in $\epsilon$ can be attributed to a dynamical transition
of the backbone. Indeed, the crossover in the decay of the kinetic energy 
reported in Fig.~\ref{fig1} suggests that the backbone
behaves as a rigid structure only for $\tau > \tau_c$.
If this is the case, for $\tau > \tau_c$ rattlers move in rigid cages, and 
dissipate in each collision an amount of energy equal to $\Delta K =(1-e) K$,
with $e$ the restitution coefficient, in agreement with the $\epsilon = 0$ expectation.
Conversely, for $\tau < \tau_c$ rattlers are able to transfer energy to the rigid structure, 
so that the energy lost in a collision is larger than $(1-e) E$, which leads to a
positive $\epsilon$. Summarizing, the long--time regime, $\psi=\psi_2$, is observed 
when cages become rigid, whereas the first short--time regime, $\psi=\psi_1$, 
reflects the relaxation of cages towards their stationary configuration.  
In this picture, the inelastic collisions with the rattlers
are the main dissipation mechanisms leading to the freezing of the backbone.
Thus, the time $\tau_c$ is expected to be inversely proportional to the dissipated
energy.

This hypothesis leads to a relation between $\tau_c$ and the constant $A$ in Eq.~\ref{eq1}. 
Indeed, considering that the energy dissipated in the interval $[t_0,\tau_c)$ scales as  $A^{3/2+\epsilon}$
and that $\psi_1=\frac{2}{1+2\epsilon}$, 
one has
\be
K(\tau) \propto \tau_c^{-\alpha} G \left( \frac{\tau}{\tau_c} \right ), 
\label{eq2}
\ee
with $\alpha=\psi_1+\frac{\psi_1}{1+\psi_1}$ and $G(x)=x^{-\psi_1}(x+1)^{-\psi_2+\psi_1}$.

\begin{figure}[!t]
\begin{center}
\includegraphics[scale=0.38]{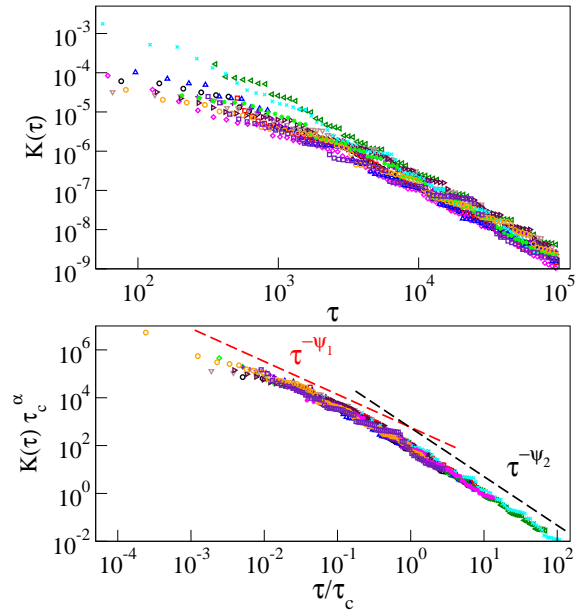}
\end{center}
\caption{\label{fig:slips}
(Color online)
(Upper panel) Decay of the kinetic energy of undriven replicas
performed at different times, following different slips occurring at time $t_s$,
as a function of the time $\tau=t-t_s$. Each color and symbol refers to a different replica of the system.
(Lower panel) Data of upper panel  collapse when time is rescaled by the crossover time,
$\tau_c$, and energy by $\tau_c^{-\alpha}$, as predicted by Eq.~\ref{eq2}.}
\end{figure}
To verify this prediction, we have investigated the relaxation of replicas created after slips
of different size. As illustrated in Fig.~\ref{fig:slips},
the raw data for the relaxation dynamics of the kinetic energy (upper panel) nicely collapse when the 
energy and time are rescaled
according to Eq.~\ref{eq2} (lower panel), confirming our interpretation.

\begin{figure}[t!!!]
\begin{center}
\includegraphics[scale=0.3]{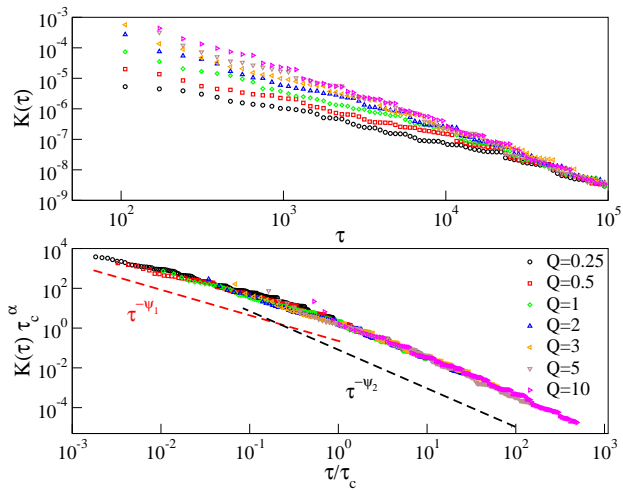}
\end{center}
\caption{(color online) 
(Upper panel) Relaxation of the kinetic energy $K(\tau)$ of different replicas, following a rescaling of the rattlers velocity by a factor $Q$.
(Lower panel) Data collapse  according to Eq.~\ref{eq2}.
\label{figQ}
}
\end{figure}

In addition, we have performed simulations to better clarify the role of the rattlers in the stabilization of the backbone.
Indeed, if the backbone dissipates energy and freezes mainly interacting with the rattlers, then the crossover time
$\tau_c$ is expected to decrease on increasing the backbone-rattlers collision frequency. 
Thus, we have monitored the relaxation dynamics of
the replicas, after scaling the velocity of each rattler
by a factor $Q$, i.e. $v \to Q v$, keeping $Q$ small enough not to destabilize the backbone.
Results, plotted in Fig.~\ref{figQ} for a given replica and for $Q \in [0.25,10]$, show that $\tau_c$ 
decreases with $Q$, whereas $A$ increases with $Q$. 
In Fig.~\ref{figQ} (lower panel) we plot $\tau_c^{\alpha} K$ as function of $\tau/\tau_c$
obtaining a good data collapse for all values of $Q$ in agreement with the scaling relation Eq.~\ref{eq2}.
Analogous results are obtained for different replicas.

\begin{figure}[!!t]
\begin{center}
\includegraphics*[scale=0.3]{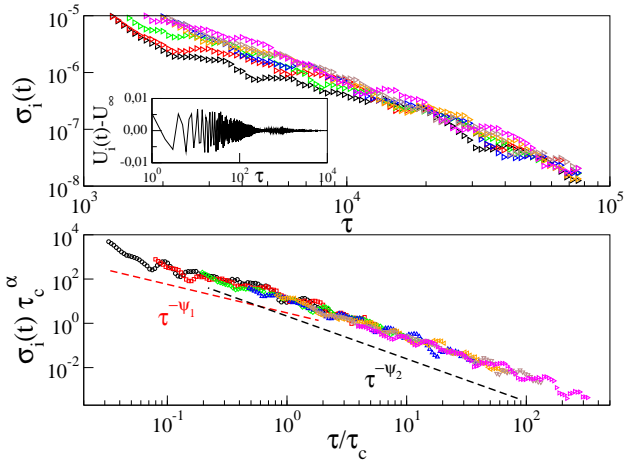}
\end{center}
\caption{(Color online) (Upper panel) The standard deviation $\sigma_{i}(t)$, evaluated in logarithmic binned intervals, is plotted versus $\tau$ for different values of $Q$. We adopt the same symbols and colors of Fig. 5. In the inset the evolution of the potential energy for $Q = 1$. (Lower panel) The same data of $\sigma_{i}(t)$  rescaled according to Eq. (2) with the same value of $\alpha$ and $\tau_{c}$ used in Fig. 5.
} 
\label{figE}
\end{figure}

The above interpretation is supported  by the evolution of the elastic energy $U(t)$ of the backbone particles which, 
during the undriven phase, presents oscillations around a constant value $U_\infty$,
as illustrated in the inset of Fig.~\ref{figE}. 
The fluctuations of this elastic energy,
$\sigma (t) =\sqrt {\left \langle  U^2(t) \right \rangle -\langle U(t) \rangle ^2 }$, 
where $\langle \cdot  \rangle$ indicate temporal averages
over a time scale $t_{\rm ave} = 10$, are a proxy of the degree of stiffness of the backbone.
Fig.~\ref{figE} (upper panel) shows that the fluctuations of the elastic energy decays in time exhibiting a double power law, with a crossover time occurring at time
$\tau_c$, alike the kinetic energy of the rattlers $K(t)$, with exponents compatible with $\psi_1$ and $\psi_2$.
Similarly, the fluctuations of the elastic energy of the backbone, in the presence of a rescaling of the velocity of the rattlers by a factor $Q$,
follow the same scaling as the overall kinetic energy, as in Fig.~\ref{figE} (to be compared with Fig.~\ref{figQ}).
According to the fluctuation-dissipation theorem, fluctuations in the elastic energy can be related to the response of the system to an external perturbation. 
As a consequence, the energy fluctuations can be considered a probe of the stiffness of the system, and the crossover at $\tau_c$
is an indication of a change in the elastic properties of the backbone. 

\section{Discussion}
Investigating the dynamics of a granular system in the jammed state, we have shown
that when the system jams due to the emergence of a network of contact forces able to sustain
the applied stress, particle motion is not suppressed. 
Rather, the overall kinetic energy of the system becomes dominated by few rattler particles,
that scatter inside their cages. Rattler-backbone collisions, surprisingly,
stabilize the backbone. Thus, the larger the kinetic energy of the rattlers when
the system jams, the smaller the time needed by the backbone to become rigid.

As a final remark we stress that there are strong similarities between the 
energy relaxation observed in our system and the one detected in the free 
cooling granular gas~\cite{Haff,pathak_FCGG}. 
Notwithstanding the striking difference in the granular 
density between the two systems, both exhibit a double power law decay with 
very similar values of the two exponents $\psi_1$ and $\psi_2$. 
Exploring if this similarity reflects some common mechanism beyond the relaxation of the two 
systems or it is just a coincidence is an interesting point to be addressed in 
future studies. 
Conversely, the relaxation of sheared frictionless granular systems
occurring when the shear is turned off is qualitatively different, since it is 
found to follow an exponential decay~\cite{Olsson2015}.
%\bibliography{granular}{}
%\input{slowrelax9.bbl}

\end{document}